\documentclass[11pt,a4paper]{article}
\usepackage[hyperref]{bert}
\usepackage{times}
\usepackage{latexsym}
\usepackage{url}
\usepackage{graphicx}
\usepackage{float}

\aclfinalcopy 

\title{End-to-End Resume Parsing and Finding Candidates for a Job Description using BERT}

\author{Vedant Bhatia 
\\IIIT Delhi
\\\textit{vedant16113@iiitd.ac.in}
\\\And
Prateek Rawat 
\\IIIT Delhi
\\\textit{prateek17041@iiitd.ac.in}
\\\And
Ajit Kumar
\\Adobe
\\\textit{ajikumar@adobe.com}
\\\AND
Rajiv Ratn Shah 
\\MIDAS, IIIT Delhi
\\\textit{rajivratn@iiitd.ac.in}
\\}

\date{}

\begin{document}

\maketitle
\begin{abstract}
  The ever-increasing number of applications to job positions presents a challenge for employers to find suitable candidates manually. We present an end-to-end solution for ranking candidates based on their suitability to a job description. We accomplish this in two stages. First, we build a resume parser which extracts complete information from candidate resumes. This parser is made available to the public in the form of a web application. Second, we use BERT sentence pair classification to perform ranking based on their suitability to the job description. To approximate the job description, we use the description of past job experiences by a candidate as mentioned in his resume. Our dataset comprises resumes in LinkedIn format and general non-LinkedIn formats. We parse the LinkedIn resumes with 100\% accuracy and establish a strong baseline of 73\% accuracy for candidate suitability. 
\end{abstract}



\section{Introduction}

Nowadays employers receive hundreds of applications for each open position. It is not possible to manually evaluate each and every application, in fact, it can be a waste of human resources to try doing so. However, the task of determining whether an applicant is suitable for a certain job profile requires human intelligence. There are a multitude of factors which cannot be automated yet, such as evaluation of the candidate's soft skills, accounting for company demands (which vary from time to time), and judging the veracity of the candidate's resume. However, we can automate a part of the process and lower the number of candidates which a human needs to judge. A vital part of the hiring process is to read the candidate's resume and judge whether they are suitable for a particular job description or not. This task is not difficult for a human. Reading and assimilating information from a resume, and comparing it with a job description to judge the suitability of the application is a task most literate humans can perform. Therefore, this task is immensely difficult for a machine to perform. Traditional computer systems fail to recognize the underlying semantic meaning for different resumes. However, the recent progress in machine learning and natural language processing (NLP) techniques means that many tasks can be performed with par-human performance. We can split the task of automating resume shortlisting into two sub-tasks. The first sub-task is parsing the resume, i.e., extracting information in a structured format from the document. The second sub-task is extracting semantic information and actually understanding the underlying information. We can then use this information to perform classification or ranking or matching tasks, as a human would do. 
\\
Resumes come in myriad formats, and simply parsing the resume correctly is a very difficult task for a machine. Current approaches normally assume a standard format which they can parse, or assume that the parsed information is available to use for their task. Our dataset comprises two categories of resumes, LinkedIn format and non-LinkedIn generic format. 
We use document metadata based heuristic rules to classify a resume as either LinkedIn or non-LinkedIn.
The LinkedIn format resumes are then converted into HTML format. This allows us to extract various style properties for each portion of the resume that would have been lost if we had just converted the PDF to text. Due to the consistent format that LinkedIn resumes follow we are able to extract all the information in the PDF file in a structured manner. 
We use the LinkedIn format resumes to train a classifier which converts the non-LinkedIn format to LinkedIn format resumes for data augmentation and data uniformity (for recruiters) purposes. We also create a web application for recruiters to use so that they may parse the resumes and obtain the candidate information in a manipulable format. Our parser extracts 100\% information with no loss of structure from LinkedIn resumes. We have also explored the feasibility of building a resume parser which can parse any resume regardless of the format.

We have created a deep-learning based system which  ranks applicants for job positions based on their suitability to the job description. To achieve this, we use the state of the art language representation model, BERT \cite{devlin2018bert}. BERT has set the state-of-the-art in a large number of NLP tasks. It pre-trains deep bidirectional representations by jointly conditioning on both left and right context in all layers. As a result, the pre-trained BERT representations can be fine-tuned with just one additional output layer to create state-of-the-art models for a particular task without substantial task-specific architecture modifications. We use BERT for sequence classification to classify the segments of text from non-LinkedIn resumes to LinkedIn format sections (Bio, Experience, Education, etc.). We perform job description and candidate suitability ranking by exploiting BERT sequence pair classification of the work experience of the candidates and the job description and using the score as a degree of suitability, allowing us to rank the resumes. In order to simulate domain specific job descriptions we use a candidate's past job experience description as a real job descriptions, as they detail their responsibilities at the job, which is what the job descriptions also contain. Through this method, we establish a strong baseline for candidate-job description suitability ranking.
The layout of the paper is as follows. We discuss related work in section \ref{related} and our system overview in section \ref{method}. The dataset (section \ref{dataset}), results (section \ref{results}) and conclusion (section \ref{conclusion}) follow. 
Our contributions through this paper are:
\begin{enumerate}
    \item Exploring the feasibility of building a generic resume parser which performs across different formats
    \item Building a heuristic and BERT based model to convert resumes to a standard format
    \item Building an end-to-end resume parsing and data collection tool for employers in the form of a web application
    \item Ranking resumes as per their suitability to a job description using BERT sequence pair classification
    \end{enumerate}

\begin{figure}[t]
    \centering
    \includegraphics[width=0.44\textwidth]{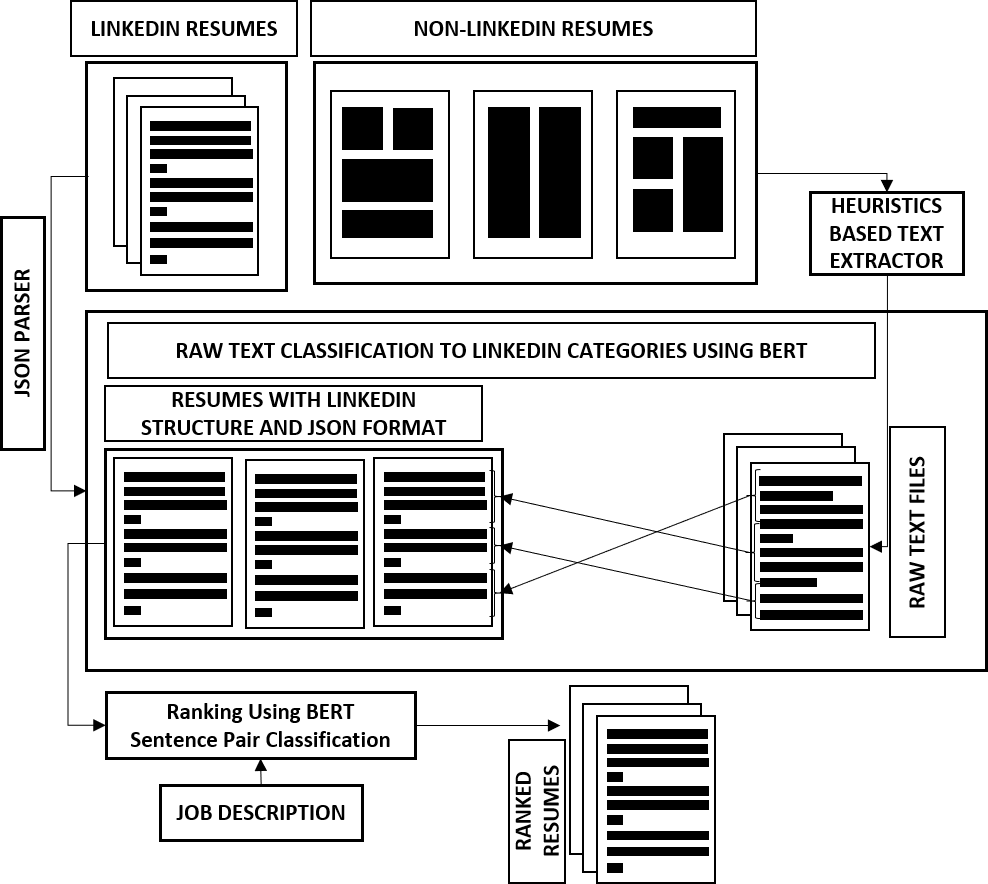}
    \caption{System diagram illustrating the data movement and tasks performed}
    \label{fig:datamovement}
\end{figure}

\section{Related Work} \label{related}
The state of the art tools in information extraction from resumes and hiring process automation are privatized by companies, both as in-house tools and commercial products. As such, although the machine learning era should have resulted in great progress in this field, the best of work done has perhaps not been seen by the research community. The parser that we have made will be made available to the public as a web application.
Most hiring process automation work does not include extracting information from PDFs \cite{kumaran2013towards,yi2007matching,al2012survey}. \cite{chen2016information} focused specifically on information extraction from PDF documents, by first classifying the document into blocks using heuristic methods and then using conditional random fields to label the different blocks. We also follow heuristic methods for classification into blocks for non-LinkedIn resumes, but we then use BERT sequence classification to classify the blocks, which is more powerful technique. Most work \cite{chen2016information,yu2005resume,singh2010prospect,chen2018two} did not extract all the information from the resume, instead focusing on certain portions like personal information and education sections. \cite{lin2016machine} extracted manual and cluster features, training a Chinese word2vec \cite{mikolov2013efficient} and concluding that learning based methods perform better for this task than manual rule based methods. However, embedding models word2vec \cite{mikolov2013efficient} and GloVe \cite{pennington2014glove} have been shown to be less effective for documents \cite{le2014distributed}. The performance also varies from section to section \cite{singh2010prospect}. Previous work has also seen ontology based experiments \cite{kumaran2013towards}, however these have largely been conceptual works. Structured relevance models have also been used to match resumes and jobs \cite{yu2005resume}, however the results were poor, with only one in 35 relevant resumes placed in the top 5 predicted candidates for a job description. \cite{maheshwary2018matching} uses a deep Siamese network \cite{bromley1994signature,chopra2005learning} to match resumes to jobs descriptions. The data collected include nearly three times as many job-descriptions as resumes, and there was no domain restriction for the job descriptions whereas the resumes were all collected from applications to one research position. They also do not deal with information extraction from PDFs.

\section{Method} \label{method}

\subsection{Building a Parser for Generic Resumes} \label{4.1}

Resumes typically have no fixed format. They are semi-structured documents and the layout is entirely up to the creator. Some common formats include a list format, a tabular format, dual-list format, or an unordered blob-based format. We can safely assume that the resume will be in PDF format, as the vast majority of employers mandate this. Different schematics whic non-LinkedIn resumes follow are shown in Figure \ref{fig:datamovement}. Ideally, a PDF created with proper tools will have metadata information such as font size and font type available for each character. However, the vast majority of resumes we came across did not have metadata information available as they were not created with proper tools. 
\\
The first step of information extraction from resumes is to extract the text from the PDF document. We use a combination of two tools to convert the PDF document to text.
\begin{itemize}
    \item \href{https://poppler.freedesktop.org/}{pdftohtml}
    \item \href{https://tika.apache.org/}{Apache Tika}
\end{itemize}

Once the raw text (with or without metadata) has been extracted from the PDF, we need to structure the raw text into categories. Humans can tell which part of the text corresponds to headings - using vision based cues such as font size, text placement, and page structure, as well as semantic information which humans naturally capture.
We developed heuristics based on the information which the tools gave us. The two main tasks involved were:
\begin{enumerate}
    \item Extracting spatially consistent text: In list style resumes, the extracted text will be spatially consistent - i.e., the logical flow of the text will match the extracted physical flow of the text. However, for other cases, such as a two-column list format, the logical flow is not the same as the physical flow (left to right across the page). To tackle this issue of the logical flow of text, we built heuristic rules based on the spacing of text across the page. Larger spaces between words on the same horizontal line would indicate inter-column gaps whereas smaller spaces would indicate normal inter-word gaps. The number of inter-word gaps also greatly exceeds the number of inter column gaps.
    Using these heuristic rules we modified the naive physical flow returned by the conversion tools to approximate the true logical flow of the document.
    \item Clustering the extracted text into different headings: Once we have the text in a logical format, we still need to cluster different parts of the text into a fixed set of headings. If the text is logically ordered, then this is a matter of identifying the different headings in the body of the text and using these to partition the text into different segments. 
\end{enumerate}
Apart from these heuristic rules, we also attempted to use vision based techniques to identify tables and section boundaries on the page, however due the immense variation in styles and formats we were not able to achieve meaningful results with this approach.

\subsection{Building a Parser for a Standard Format of Resumes}

The largest professional social network in the world is LinkedIn. Almost every professional has a LinkedIn profile with the same information as that on their resumes, and frequently applies for jobs or networks with recruiters through the website. We exploit the prevalence of LinkedIn in the professional world to use their resume format as our standard format. This format is well structured (a structured schematic is shown in Figure \ref{fig:datamovement}). Each section (such as Bio, Experience, Education, etc.) is demarcated from the others. The format within sections is also fixed, which aids information extraction. We follow a heuristic based approach to extract information. The logical flow of the document always follows the following structure:
\begin{enumerate}
    \item Personal information: Name, location, current employment, contact information etc.
    \item Career information: Summary, Experience, Education etc.
    \item Recommendations: Recommendations from other professionals
\end{enumerate}

In addition to this fixed structure, the PDF metadata is intact and consistent - the font size, style etc. are always available and the properties of headings stay the same across different resumes.

The combination of the consistent structure and the metadata information available results in very successful heuristic based information extraction from these resumes. We are always able to extract the entirety of the information in the resumes keeping the structure intact. We use pdftohtml to extract the text and metadata information, and use our heuristics to segment the text into the original sections, finally outputting a JSON or CSV file.

\subsubsection{Converting Resumes to LinkedIn Format}
In order to perform data augmentation for various LinkedIn resume based tasks as well as give employers a common format to work with, we attempted to use neural networks to convert resumes in general formats to the LinkedIn format. To accomplish this we first split the resumes into different segments based on the heuristic rules we had developed in Section \ref{4.1}. We used BERT to create feature vectors for these segments. We train a classifier on the LinkedIn format resumes, and then classify each segment of the non-LinkedIn resumes into one of the LinkedIn format segments. The results are discussed in Section \ref{results}.

\subsubsection{Web Application for Resume Parsing}

We made our resume parser for LinkedIn resumes available to requiters through a web-application. The recruiter can upload a batch of resumes and the application will return the parsed information in the form of a CSV file. The file has all the information available in the resumes for each applicant, and the recruiter can record the comments for each section of the resume as the candidate is interviewed. The CSV file also allows the comments for the different stages of the recruitment process to be recorded conveniently. Thus, it results in uniform and efficient data collection for the employers. The increased amount of well-structured, uniform data with interviewer comments can be the basis of future work in this field.

\subsection{Ranking candidates on the basis of job-description suitability}

We use the information extracted from the LinkedIn format resumes to perform this ranking. We restrict ourselves to the candidates past professional experience. A typical description of the individuals previous employment details his responsibilities at the job. 
In order to simulate the job description that a company would be looking to hire individuals for, we used one candidate's description of his responsibilities at a previous role as the job description. We created positive samples by taking combinations of different job responsibilities that a person had. For example, person \(P_1\) had job experiences \(E_{11}\), \(E_{12}\), \(E_{13}\) so the combinations \((E_{11},E_{12})\), \((E_{11},E_{13})\), \((E_{12},E_{13})\) were positive samples. Combination of job responsibilities between different candidates were taken as negative samples. For example, \(P_1\) and \(P_2\) had job experience \(E_{11}\) and \(E_{21}\) respectively, so we created a negative sample as \((E_{11},E_{21})\).
Using these we trained BERT for sequence pair classification (Figure \ref{fig:datamovement}) task to predict whether the two job descriptions were of the same candidate or not. BERT gives us a score in the range [0,1], ranging from no-match to a perfect match. 

Given that the two job descriptions of a person are not exactly same but are in the same domain, the dataset creation and training procedure we followed allowed us to learn job description similarity without having a labelled dataset. It removed the requirement of a domain expert or even multiple domain experts which could tell if a person having a particular job experience is eligible for job at hand. 
Once we had a trained model we could simply input job description and candidate's work experience and use the classification score to rank different candidates.
\section{Evaluation} \label{evaluation}
\subsection{Dataset} \label{dataset}

Our dataset comprised two parts, LinkedIn resumes and non-LinkedIn resumes. We had 715 LinkedIn resumes. Of these 305 had the work experience detailed properly, and hence we moved forward with these. We had 1,000 non-LinkedIn resumes in PDF format. 
\\
For the candidate ranking task, we required job-descriptions in order to train our model. As we did not have any data  to classify whether a job description matched the candidate experience, we created our own dataset by splitting job experience of a candidate for different companies and taking binary combinations of those as positives. For negative sampling, we randomly selected job experiences of other candidates. Doing this for each candidate resulted in 3958 samples on which we trained BERT for sentence pair classification of text segments into LinkedIn format sections (Bio, Experience, Education, etc.). 
\subsection{Results} \label{results}

Over the course of our algorithm (shown in Figure \ref{fig:datamovement}), we have performed three classification tasks and one ranking and similarity computation task.\\
The performance of our classifiers follows:
\begin{enumerate}
    \item Differentiating between LinkedIn and non-LinkedIn resumes: We use a heuristic based method to tell the difference between LinkedIn and non-LinkedIn resumes. The distribution of font  size frequency and order of occurrence of different font sizes allows us to identify a LinkedIn resume. Due to this heuristic based method, on a test set of 100 LinkedIn and 100 non-LinkedIn resumes, we achieved 100\% accuracy.
    \item \label{clf2} Structuring extracted text from LinkedIn resumes into predefined segments: This classifier classifies the text extracted from the LinkedIn format (along with the metadata) into different headings and outputs a JSON file. We use heuristic based methods here as well, identifying a heading from normal text, as well as identifying the subheadings. The uniform format of document styling allows us to perform this with perfect accuracy as well. On a test set of 100 LinkedIn resumes, we achieved classification with 100\% accuracy into the sub-categories.
    \item Converting a non-LinkedIn resume to a LinkedIn format resume: This classifier classifies segments of text into different LinkedIn categories. We use heuristic based methods to grab segments text of from the raw text extracted. We then use BERT for sequence classification, which we fine tuned on results from our second classifier (\ref{clf2}) to predict which class the segment belongs to. We achieved 97\% accuracy on a test set of 35 manually annotated resumes.
\end{enumerate}

We used BERT for sentence pair classification and achieved 72.77\% accuracy in predicting whether two job descriptions belong to same person or not. This method can be used to predict whether a person's previous job experience is similar to a job description at hand. Due to a lack of ground truth, we cannot train a network to rank resumes as per their suitability to the job description. Thus, we use the sentence pair classification score from BERT as the ranking criterion, as this intuitively gives us a degree of similarity between the job description and profiles of candidates.  
\section{Conclusion} \label{conclusion}
Through this paper we explore the feasibility of creating a standard parser for resumes of all formats. We found that this was not possible to do without information loss for all cases, which would result in the unfair loss of certain applicant's resumes in the process. We instead proceeded with LinkedIn format resumes, for which we could build a parser with no information loss. This parser is publicly available through a web application. \\We used BERT for sequence pair classification to rank candidates as per their suitability to a particular job description. With the data collected from the web application, we will have real job descriptions and interviewer comments at each stage of the hiring process. We also plan to further explore the vision based page segmentation approach in order to augment our structural understanding of resumes. This work establishes a strong baseline and a proof of concept which can lead to the hiring process benefiting from the advances in deep-learning and language representation.
\bibliographystyle{bert}
\bibliography{bert}

\begin{thebibliography}{15}
\expandafter\ifx\csname natexlab\endcsname\relax\def\natexlab#1{#1}\fi

\bibitem[{Al-Otaibi and Ykhlef(2012)}]{al2012survey}
Shaha~T Al-Otaibi and Mourad Ykhlef. 2012.
\newblock A survey of job recommender systems.
\newblock \emph{International Journal of Physical Sciences}, 7(29):5127--5142.

\bibitem[{Bromley et~al.(1994)Bromley, Guyon, LeCun, S{\"a}ckinger, and
  Shah}]{bromley1994signature}
Jane Bromley, Isabelle Guyon, Yann LeCun, Eduard S{\"a}ckinger, and Roopak
  Shah. 1994.
\newblock Signature verification using a" siamese" time delay neural network.
\newblock In \emph{Advances in neural information processing systems}, pages
  737--744.

\bibitem[{Chen et~al.(2016)Chen, Gao, and Tang}]{chen2016information}
Jiaze Chen, Liangcai Gao, and Zhi Tang. 2016.
\newblock Information extraction from resume documents in pdf format.
\newblock \emph{Electronic Imaging}, 2016(17):1--8.

\bibitem[{Chen et~al.(2018)Chen, Zhang, and Niu}]{chen2018two}
Jie Chen, Chunxia Zhang, and Zhendong Niu. 2018.
\newblock A two-step resume information extraction algorithm.
\newblock \emph{Mathematical Problems in Engineering}, 2018.

\bibitem[{Chopra et~al.(2005)Chopra, Hadsell, LeCun
  et~al.}]{chopra2005learning}
Sumit Chopra, Raia Hadsell, Yann LeCun, et~al. 2005.
\newblock Learning a similarity metric discriminatively, with application to
  face verification.
\newblock In \emph{CVPR (1)}, pages 539--546.

\bibitem[{Devlin et~al.(2018)Devlin, Chang, Lee, and
  Toutanova}]{devlin2018bert}
Jacob Devlin, Ming-Wei Chang, Kenton Lee, and Kristina Toutanova. 2018.
\newblock Bert: Pre-training of deep bidirectional transformers for language
  understanding.
\newblock \emph{arXiv preprint arXiv:1810.04805}.

\bibitem[{Kumaran and Sankar(2013)}]{kumaran2013towards}
V~Senthil Kumaran and A~Sankar. 2013.
\newblock Towards an automated system for intelligent screening of candidates
  for recruitment using ontology mapping (expert).
\newblock \emph{International Journal of Metadata, Semantics and Ontologies},
  8(1):56--64.

\bibitem[{Le and Mikolov(2014)}]{le2014distributed}
Quoc Le and Tomas Mikolov. 2014.
\newblock Distributed representations of sentences and documents.
\newblock In \emph{International conference on machine learning}, pages
  1188--1196.

\bibitem[{Lin et~al.(2016)Lin, Lei, Addo, and Li}]{lin2016machine}
Yiou Lin, Hang Lei, Prince~Clement Addo, and Xiaoyu Li. 2016.
\newblock Machine learned resume-job matching solution.
\newblock \emph{arXiv preprint arXiv:1607.07657}.

\bibitem[{Maheshwary and Misra(2018)}]{maheshwary2018matching}
Saket Maheshwary and Hemant Misra. 2018.
\newblock Matching resumes to jobs via deep siamese network.
\newblock In \emph{Companion of the The Web Conference 2018 on The Web
  Conference 2018}, pages 87--88. International World Wide Web Conferences
  Steering Committee.

\bibitem[{Mikolov et~al.(2013)Mikolov, Chen, Corrado, and
  Dean}]{mikolov2013efficient}
Tomas Mikolov, Kai Chen, Greg Corrado, and Jeffrey Dean. 2013.
\newblock Efficient estimation of word representations in vector space.
\newblock \emph{arXiv preprint arXiv:1301.3781}.

\bibitem[{Pennington et~al.(2014)Pennington, Socher, and
  Manning}]{pennington2014glove}
Jeffrey Pennington, Richard Socher, and Christopher Manning. 2014.
\newblock Glove: Global vectors for word representation.
\newblock In \emph{Proceedings of the 2014 conference on empirical methods in
  natural language processing (EMNLP)}, pages 1532--1543.

\bibitem[{Singh et~al.(2010)Singh, Rose, Visweswariah, Chenthamarakshan, and
  Kambhatla}]{singh2010prospect}
Amit Singh, Catherine Rose, Karthik Visweswariah, Vijil Chenthamarakshan, and
  Nandakishore Kambhatla. 2010.
\newblock Prospect: a system for screening candidates for recruitment.
\newblock In \emph{Proceedings of the 19th ACM international conference on
  Information and knowledge management}, pages 659--668. ACM.

\bibitem[{Yi et~al.(2007)Yi, Allan, and Croft}]{yi2007matching}
Xing Yi, James Allan, and W~Bruce Croft. 2007.
\newblock Matching resumes and jobs based on relevance models.
\newblock In \emph{Proceedings of the 30th annual international ACM SIGIR
  conference on Research and development in information retrieval}, pages
  809--810. ACM.

\bibitem[{Yu et~al.(2005)Yu, Guan, and Zhou}]{yu2005resume}
Kun Yu, Gang Guan, and Ming Zhou. 2005.
\newblock Resume information extraction with cascaded hybrid model.
\newblock In \emph{Proceedings of the 43rd annual meeting on association for
  computational linguistics}, pages 499--506. Association for Computational
  Linguistics.

\end{thebibliography}

\end{document}